\documentclass[aps,pre,showpacs,amsmath,amssymb,amsfonts,superscriptaddress,twocolumn]{revtex4}

\usepackage{graphicx}
\usepackage{subfigure}
\usepackage{verbatim}
\usepackage{dcolumn}
\usepackage{bm}
\usepackage{epsf}
\usepackage{color}
\usepackage[colorlinks=true,citecolor=blue,linkcolor=blue]{hyperref}

\newcommand{\bla}{bla\\bla\\bla\\bla\\bla}

\begin{document}

\title{Thermodynamics of Information Processing Based on Enzyme Kinetics: an Exactly Solvable Model of Information Pump}

\author{Yuansheng Cao}
\affiliation{School of Physics, Peking University, Beijing 100871, China}
\author{Zongping Gong}
\affiliation{School of Physics, Peking University, Beijing 100871, China}
\author{H. T. Quan}
\email{htquan@pku.edu.cn}
\affiliation{School of Physics, Peking University, Beijing 100871, China}
\affiliation{Collaborative Innovation Center of Quantum Matter, Beijing 100871, China}

\date{\today}

\begin{abstract}
Motivated by the recent proposed models of the information engine [D. Mandal and C. Jarzynski, Proc. Natl. Acad. Sci. {\bf109}, 11641 (2012)] and the information refrigerator [D. Mandal, H. T. Quan, and C. Jarzynski, Phys. Rev. Lett. {\bf111}, 030602 (2013)], we propose a minimal model of the information pump and the information eraser based on enzyme kinetics. This device can either pump molecules against the chemical potential gradient by consuming the information encoded in the bit stream or (partially) erase the information encoded in the bit stream by consuming the Gibbs free energy. The dynamics of this model is solved exactly, and the ``phase diagram" of the operation regimes is determined. The efficiency and the power of the information machine is analyzed. 
 The validity of the second law of thermodynamics within our model is clarified. Our model offers a simple paradigm for the investigating of the thermodynamics of information processing involving the chemical potential in small systems.
\end{abstract}


\maketitle

\newpage

\section{Introduction}


In 1871 \cite{Maxwell1871}, in a thought experiment, James C. Maxwell conceived an intelligent creature, now known as Maxwell's demon \cite{Leff2003} (MD), to elaborate the statistical nature of the second law of thermodynamics. In this thought experiment it was assumed that this demon can separate the fast-moving molecules from the slow ones in the gas contained in a box, thus apparently ``violate" the Clausius' formulation of the second law. Since its conception, the consequence of the MD has been under debate by a lot of researchers, including many renowned physicists for many decades \cite{Leff2003}, and the MD has become one of the most famous thought experiments in the history of physics. Many interpretations on the conceptual levels have been put forward to guarantee that the MD will not lead to the violation of the second law. Nevertheless, it was not until 1982 when C. H. Bennett gave an explanation \cite{Bennett1982} based on a principle proposed by R. Landauer \cite{Landauer1961} (and independently by O. Penrose \cite{penrose1970}), a consensus is finally reached. According to Bennett \cite{Bennett1982}, information, or Shannon's entropy, of the memory register of the demon plays a key role in preventing the second law from being violated in the Maxwell's thought experiment. The memory register of the demon has to be erased in a cycle, consuming \(k_BT\ln2\) energy per bit (Here \(k_B\) is the Boltzmann's constant, and \(T\) is the temperature of the reservoir). When the erasure of the memory register is included into the cycle, the total entropy production can never be negative. Thus the second law of thermodynamics will not be violated by the MD \cite{Bennett1982}.


In principle, the low-entropy state of the memory register can help rectify the thermal fluctuations, and convert one form of disordered energy (e.g., the heat) into another form of ordered energy (e.g., the work). The cost is that the Shannon's entropy of the memory register will increase by at least the amount of the entropy decrease in the non-information-bearing degrees of freedom (e.g. the heat reservoir). In practice, however, the implementation of the Maxwell's thought experiment in a laboratory and the realization of the ``tradeoff" between the energy and the information has not been fully demonstrated in the laboratory. It has attracted increased attention in recent years to bring Maxwell's thought experiment into practice in either artificial or natural systems. The candidates include systems borrowed from the quantum information processing \cite{chuan2012quantum,jacobs2012quantum,quan2006}, the signal transduction in biological network \cite{Tu-2,sartori2014thermodynamic}, the AMO systems \cite{Raizen2006,Binder2008,Raizen2009,Raizen2011}, and the artificial nano-scale devices with feedback control \cite{serreli2007,kim2007fluctuation,Jacobs2009,Sagawa2010,toyabe2010experimental,ponmurugan2010,horowitz2010,Abreu2011,abreu2012,Suri2011,dong2011quantum,strasberg2013,Deffner2013,Jordan2014a,Barato2014,Parrondo2015}. In almost all these models, an intelligent being or an external controller is involved (one exception is \cite{Raizen2011}). It is less interesting to have an intelligent being or an external controller involved when realizing Maxwell's thought experiment since there is no demon in the world. 
Thus, it 
has become a goal for many physicists to build an ``autonomous" information machine, which can rectify the thermal fluctuations and mimic what Maxwell has conceived but without involving any intelligent being.
Along this line, an autonomous information engine model \cite{Mandal2012,lu2014engineering} and an autonomous information refrigerator model \cite{Mandal2013} have been proposed. 
Also, an optical MD, which can reduce the entropy of the atoms at the expense of increasing the entropy of scattered photons has been explored \cite{Raizen2006,Binder2008,Raizen2009,Raizen2011}. Nevertheless it is difficult to characterize the entropy increase in the scattered photons and study the second law as well as the efficiency and the power of the optical MD.


In this paper we propose a minimal solvable model of an autonomous information pump and eraser that connects the information and the chemical potential: it pumps molecules against the chemical potential gradient by simultaneously writing information to the memory register, without the expenditure of  work or free energy, and without the involvement of any intelligent being. Namely, it is an autonomous information machine. We notice that there are realistic molecule pumps in biological systems, such as ATPase \cite{ToyabeOkamoto-6,NojiYasuda-7,lingrel1994na+} or ion channels \cite{hille2001ion}. Usually one kind of molecules (such as  $Na^+$) are pumped against the chemical potential gradient at the cost of dissipating the free energy of other molecules (such as ATP). However, in our model, instead of consuming the chemical potential difference of another kind of molecules, we consume the information. The model is constructed based on classical enzyme reactions with three states \cite{Seifert2011,qian1997,Qian-1}, coupled to a memory register, which carries bits of either 1 or 0 in a stream (Fig. \ref{fig:1}(b)). The enzyme undergoes stochastic evolutions, which catalyses the reactions and overrides the states of the bits. In certain regions of the parameter space, the information machine evolves to a periodic steady state and continuously pumps molecules, while in other regions of the parameter space, it (partially) erases information encoded in the memory register at the cost of dissipating free energy.


This paper is organized as follows. In section II we introduce the three-state enzymatic system and the bit stream that couples to the three-state system. In section III we solve the dynamics exactly, and discuss the the efficiency and the power of our model. The validity of the second law of thermodynamics within our model is clarified. In section IV, we discuss and conclude our paper.

\section{The model}
\begin{figure*}[tbp]
\centering
\includegraphics{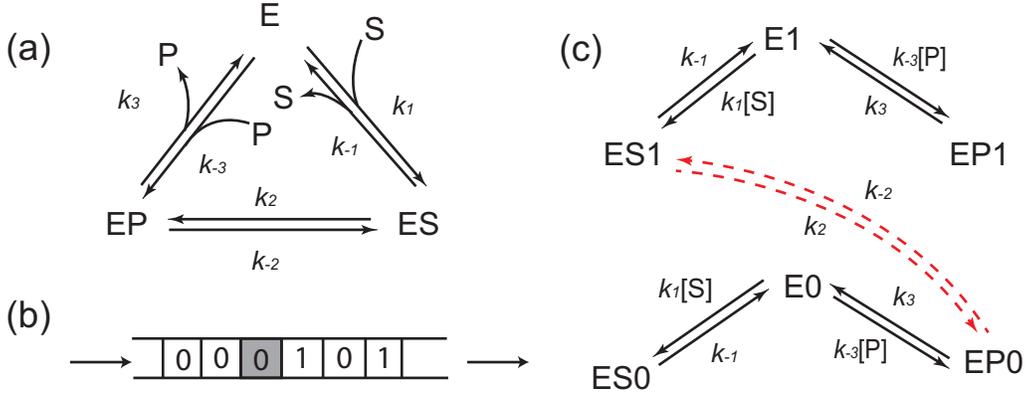}
\caption{Schematic figure of enzyme kinetics (a), the bit stream (b) and the information pump or information eraser based on the enzyme kinetics (c).}
\label{fig:1}
\end{figure*}

The original enzymatic model which describes how an enzyme {\it E} catalyzes the reaction \cite{ge2008,ge2012,Qian-1,qian1997,Seifert2011} is shown in Fig. \ref{fig:1}(a). A {\it E} molecule binds with a {\it S} molecule and forms a complex {\it ES} molecule. A {\it ES} molecule undergoes a catalytic process and becomes a new complex {\it EP} molecule. The {\it EP} molecule then releases the product -- a {\it P} molecule and the enzyme {\it E} is set free and goes into the next cycle. All these reactions take place under isothermal and isobaric conditions, so the Gibbs free energy (chemical potential) can be used to determine the direction of the reaction. The stochastic dynamics is governed by the master equation
\begin{equation}
\frac{d {\mathbf{q}}(t)}{dt}=\mathcal{R}_{0}{\mathbf{q}}(t),
\label{masterequation1}
\end{equation}
where
\begin{equation}
\mathcal{R}_{0}=\left(\begin{array}{ccc}-(k_{-1}+k_{2})&k_1[S]&k_{-2}\\k_{-1}&-(k_1[S]+k_{-3}[P])&k_3\\k_{2}&k_{-3}[P])&-(k_{-2}+k_3)\end{array}\right),
\label{threestate}
\end{equation}
and ${\mathbf{q}}(t)=[q_{ES}(t),q_{E}(t),q_{EP}(t)]^{T}$ describes the probability distribution on the three states {\it ES}, {\it E} and {\it EP}.
When the chemical potential difference between {\it P} molecules and {\it S} molecules
\begin{equation}\label{eq:1}
 \Delta \mu=-\frac{1}{\beta}\ln(\frac{K^{eq}[S]}{[P]})
\end{equation}
is fixed at a nonzero value, in the long time limit, the system reaches a nonequilibrium steady state. Here \(\beta = 1/k_B T\). \(K^{eq}=(k_1k_2k_3)/(k_{-1}k_{-2}k_{-3})\); \(k_i\) (\(i=1,2,3,-1,-2,-3\)) describes the reaction rate constant. \([P]\) and \([S]\) depict the concentration of {\it P} molecules and {\it S} molecules. In this case, the Gibbs free energy is continuously  dissipated into heat and released to the heat reservoir. The difference of the standard Gibbs free energy of formation \(\Delta G_i^{0}\) of the two states connected by the {\it i}th reaction is determined by the dynamic parameters:
\begin{equation}
\label{eq:2}\Delta G_i^{0}=-\frac{1}{\beta}\ln \left(\frac{k_i}{k_{-i}} \right) .
\end{equation}
Usually the chemical potential of {\it S} molecules should be higher than that of {\it P} molecules so that the substrate can be converted into the product. The chemical potential difference is the driving force of the transformation from {\it S} molecules to {\it P} molecules. When the chemical potential difference disappears (\(\Delta \mu=0\)), the system approaches an equilibrium state. That means the forward and the backward reactions are balanced, and there is no conversion from {\it S} molecules to {\it P} molecules, or vice versa.

We now combine this enzymatic model with the information encoded in the incoming bit stream (Fig. \ref{fig:1} (b)) and introduce the competition between the chemical potential difference $\Delta \mu$ and the information (Fig. \ref{fig:1}(c)). Our model is very similar to the information engine model proposed in Ref.~\cite{Mandal2012} and the information refrigerator model proposed in \cite{Mandal2013}. The information is encoded in the bit stream that passes through and interacts with the three-state enzyme system. A single bit in the stream, labelled with 0 or 1, can couple to the three-state system, and forms a composite system of six states, {\it E0}, {\it ES0}, {\it EP0}, {\it E1}, {\it ES1} and {\it EP1}. The information machine is based on the six-state system and can jump within $\it {EP1} \leftrightharpoons \it {E1}$, $\it {E1} \leftrightharpoons \it {ES1}$, $\it {EP0} \leftrightharpoons \it {E0}$ and $\it {E0} \leftrightharpoons \it {ES0}$ (black solid line of Fig. \ref{fig:1}(c)) without flipping the bit. This kind of evolution is simply the binding-unbinding process without any molecular conformational change. In addition, the enzyme makes a catalytic reaction from $\it {ES}$ molecules to $\it {EP}$ molecules (from $\it {EP}$ molecules to $\it {ES}$ molecules) if and only if the bit flips from 1 to 0 (from 0 to 1), as indicated by the transformation (red dashed line in Fig. \ref{fig:1}(c)) between $\it {ES1} \leftrightharpoons \it {EP0}$. Notice that there is no transformation between {\it ES0} and {\it EP1}, {\it ES1} and {\it EP1}, or between {\it ES0} and {\it EP0}. We will see that this asymmetry in the dynamics is essential for the pumping of molecules or the erasure of information. This new enzyme kinetics is governed by the master equation
\begin{equation}
\frac{d {\mathbf{p}}(t)}{dt}=\mathcal{R} {\mathbf{p}}(t),
\label{masterequation}
\end{equation}
where
\begin{widetext}
\begin{equation}
\mathcal{R}=\left(\begin{array}{cccccc}-k_{-1}&k_1[S]&0&0&0&0\\k_{-1}&-(k_1[S]+k_{-3}[P])&k_3&0&0&0\\0&k_{-3}[P]&-(k_3+k_{-2})&k_2&0&0\\0&0&k_{-2}&-(k_{2}+k_{-1})&k_1[S]&0\\0&0&0&k_{-1}&-(k_1[S]+k_{-3}[P])&k_3\\0&0&0&0&k_{-3}[P]&-k_3\end{array}\right),
\label{sixstate}
\end{equation}
\end{widetext}
and ${\mathbf{p}}(t)=[p_{ES0}(t),p_{E0}(t),p_{EP0}(t),p_{ES1}(t),p_{E1}(t),p_{EP1}(t)]^{T}$ denotes the probability distribution in the six states {\it ES0}, {\it E0}, {\it EP0}, {\it ES1}, {\it E1} and {\it EP1}. All the parameters in Eq.~(\ref{sixstate}) are the same as those in Eq. (\ref{threestate}). The difference between the six-state dynamics (\ref{masterequation}) and the three-state dynamics (\ref{masterequation1}) is that the six-state system will reach a periodic steady state, instead of a steady state, because of the interruptions of the incoming bits and the finiteness of the interaction time with every bit.
As the bit stream passes by, the three-state system is assumed to interact with only one bit (the nearest bit) at a time. The duration of the interaction with each bit is set to be \(\tau\), which is independent of other parameters and can be tuned. We assume that the incoming bit contains a mixture of 1's and 0's with probabilities \(p_1\) and \(p_0\). Let
\begin{equation}
\label{eq:3}
\delta=p_0-p_1
\end{equation}
denote the excess of 0's in the incoming bit stream. The incoming bits are statistically independent of one another. When a fresh bit comes in, it evolves with the three-state system for a time period of \(\tau\), jumping among the six states. After each interaction period, the bit in the stream can be overridden due to the dynamic evolution (\ref{masterequation}) of the composite system. At the end of the period, the final state is decomposed into the bit and the three-state system. The final state of the bit is preserved in the outgoing bit stream, and the final state of the three-state system is recycled as the initial state for the next period. For example, if the state at the end of the \(n\)th period is {\it ES1}, and the (\(n+1\))th incoming bit is 0, then the initial state of the six-state system in the (\(n+1\))th period is {\it ES0}. This effective transition does not imply an actual change in the state of a given bit, but simply reflects the replacement of an outgoing bit in state 1 with an incoming bit in state 0 \cite{Mandal2012,Mandal2013}. Given all the dynamic parameters \(k_{\pm i}\) and \(\delta\), after many periods, the three-state system will reach a periodic steady state. Finally, let \(b_n\) and \(b'_n\) denote the incoming and the outgoing state of the \(n\)th bit in the stream. The change of the bit during the \(n\)th interaction period is
\begin{equation}
\label{eq:4}
\Delta\chi_n=b'_n-b_n.
\end{equation}
Notice that \(\Delta\chi_n\) also indicates the net number of {\it P} molecules being transformed to {\it S} molecules during the \(n\)th interaction period.

Before going to the exact solutions of the dynamics, we give an intuitive explanation of how the information machine pumps molecules from the low chemical potential reservoir ({\it P} constituents in the solution) to the high chemical potential reservoir ({\it S} constituents in the solution). For this purpose, we assume all the bits in the incoming stream are in state 0, so that \(\delta=1\). In this circumstance, in every period the information machine starts from one of three states {\it E0}, {\it EP0} or {\it ES0}. The composite system of the three-state system and the bit evolves for a time period of \(\tau\), according to the dynamics described by the master equation (\ref{masterequation}). The six-state system might jump forward and backward between {\it ES1} and {\it EP0} repeatedly. At the end of the interaction period, if the six-state system ends up in state {\it E0}, {\it EP0} or {\it ES0}, it must be the case that every transition \( ES1\rightarrow EP0\) is balanced by a transition \( EP0\rightarrow ES1\); hence \(\Delta\chi_n=0\) (\ref{eq:4}) and there is no net transformation between {\it P} molecules and {\it S} molecules. If the final state is {\it E1}, {\it EP1} or {\it ES1}, we know that there is a net transition from {\it EP0} to {\it ES1}; then \(\Delta\chi_n=1\) (\ref{eq:4}), and the result is that one {\it P} molecule is transformed to one {\it S} molecule. This net transformation will be recorded in the outgoing bit stream, that a bit initially in state 0 will be finally rewritten to 1. In the long run,
\begin{equation}
\chi(t)=\sum_{n=1}^{[t/\tau]}\Delta\chi_n
\label{chi}
\end{equation}
grows with time and we will get a net transformation from {\it P} molecules to {\it S} molecules, which is proportional to $\chi(t)$, the 1's appearance in the outgoing bit stream. During this process, Gibbs free energy is generated at the cost of increasing the entropy of the bit stream, instead of the expenditure of work or free energy. Please notice that the mode of the energy conversion in the information pump is different from that of the information engine or the information refrigerator. In the information engine (refrigerator), work is generated (heat is pumped) at the cost of increasing the entropy of the bit stream.

A more general case is when the incoming bit stream is a mixture of 0's and 1's. The excess of 0 over 1 (\(\delta>0\)), or excess of 1 over 0 (\(\delta<0\)), resists or enhances the normal flux stream from {\it S} molecules to {\it P} molecules. The fraction of 0's in the incoming bit stream will affect the  flow of molecules being catalysed, as well as the bit stream that passes by. In the following we will give a quantitative analysis on these two effects.



\section{System as information pump or information eraser}
\subsection{solution to the model}

Similar to the information engine model \cite{Mandal2012} and the information refrigerator model \cite{Mandal2013}, the evolution of the composite six-state system in every period can be separated into two stages: (\romannumeral1) the dynamic evolution under the governing of Eq.~(\ref{masterequation}), and (\romannumeral2) the projection process at the end of every period, which eliminates the correlations between the three-state system and the bit. For a finite period $\tau$, the three-state system will reach a periodic steady state after many periods (this result is guaranteed by the Perron-Frobenius theorem \cite{Mandal2012}). In order to obtain the result of the periodic steady state, we follow the same procedure as that in Refs.~\cite{Mandal2012,Mandal2013}. When the three-state system reaches the periodic steady state, we use \(p'_0\) and \(p'_1\) to denote the outgoing bits' fraction for 0's and 1's. Let \(\delta'=p'_0-p'_1\) be the excess of 0's in outgoing bit stream. The average production of 1's per interaction period is characterized by
\begin{equation}
\Phi\equiv\langle\Delta\chi\rangle=p'_1-p_1=\frac{(\delta-\delta')}{2}.
\label{eq:5}
\end{equation}
As described before, for each flipping of \(0\rightarrow1\) there must be an accompanied molecular conformational transformation of the three-state system from {\it EP} to {\it ES}, and accordingly a transformation from {\it P} molecules to {\it S} molecules. The average Gibbs free energy gained by the information machine per interaction period is
\begin{equation}
\Delta G = -\Phi \Delta\mu.
\label{eq:6}
\end{equation}
A positive \(\Delta G\) indicates that the molecules flow against the chemical potential gradient, and {\it P} molecules are converted to {\it S} molecules continuously.

We use the Shannon's information entropy
\begin{equation}
\begin{split}
I(\delta)&=-p_0\log_{2}{p_0}-p_1\log_{2}{p_1}\\
&=-\frac{1-\delta}{2}\log_{2}{\frac{1-\delta}{2}}-\frac{1+\delta}{2}\log_{2}{\frac{1+\delta}{2}}
\end{split}
\label{eq:7}
\end{equation}
to characterize the information content of every bit in the incoming bit stream. Information content of every bit in the outgoing bit stream \(I(\delta')\) could be defined similarly. The information entropy increase of every bit is given by the difference of the information entropy between the outgoing and the incoming bits
\begin{equation}\label{eq:8}
\Delta I\equiv I(\delta')-I(\delta)=I(\delta-2\Phi)-I(\delta).
\end{equation}

For simplicity, we consider the case where the Gibbs free energy difference comes only from the concentration difference between \(S\) molecules and \(P\) molecules. That is we choose \(k_{\pm i }=1\), (\(\Delta G_i^0=0\)) and let \([S]=1+\epsilon\), \([P]=1-\epsilon\), \(\epsilon\in(-1,1)\). Given a set of parameters (\(\epsilon,\delta;\tau\)), we obtain the exact solution of the periodic steady state (see Appendix A) and find the expression of the current
\begin{subequations}
\begin{equation}
\Phi (\delta,\epsilon;\tau)=\frac{\delta-\epsilon}{2}[1-\frac{1}{3}K(\tau)],
\label{eq:9}
\end{equation}
where
\begin{equation}
K(\tau)=e^{-2\tau}\frac{(1+8\alpha+4\sqrt{3}\beta)-(2+7\alpha+4\sqrt{3}\beta)e^{-2\tau}}{3-(2+\alpha)e^{-2\tau}},
\label{eq:10}
\end{equation}
\end{subequations}
and \(\alpha=\cosh{(\sqrt{3}\tau)}\), \(\beta=\sinh{(\sqrt{3}\tau)}\). Notice that \(\Phi\) determines both \(\Delta I\) and \(\Delta G\). So we obtain the entropy increase in the bit stream and the free energy gaining in every period from Eq.~(\ref{eq:9}-\ref{eq:10}). The term \(\delta-\epsilon\) in Eq.~(\ref{eq:9}) demonstrates the competition between the chemical potential gradient and the information. Actually, we have got exact solutions for a general class of models, of which our current model and the model in Ref.~\cite{Mandal2012} are two special cases (see Appendix B).

\begin{figure}
\centering
\includegraphics[width=.45\textwidth]{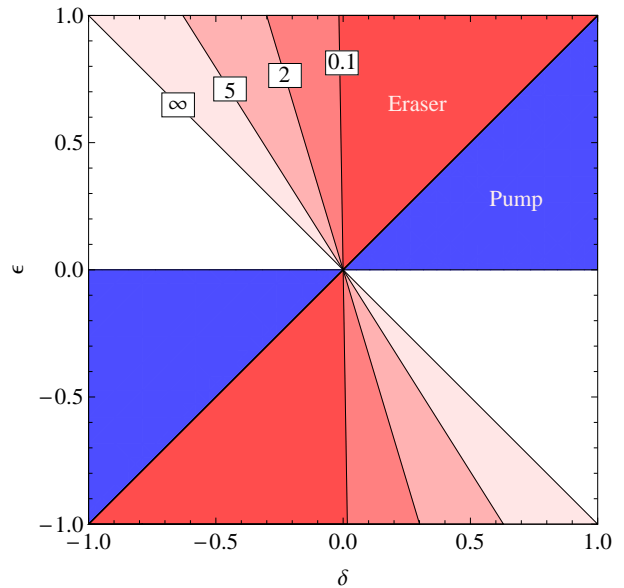}
\caption{(Color online) ``Phase diagram" of the information pump and eraser in the parameter space. The numbers indicate different lengths of the time periods $\tau=0.1,2,5,\infty$.}
\label{fig:2}
\end{figure}

In the parameter space of \(\delta\) and \(\epsilon\), we plot a ``phase diagram" (for different \(\tau\) the boundary of the ``phase" are different). The information machine operates as an information pump when the information is the predominant driving force \(0<\epsilon<\delta\), \(\Phi>0\) (more \(S\) molecules and less \(P\) molecules) or \(\delta<\epsilon<0\), \(\Phi<0\) (more \(P\) molecules and less \(S\) molecules). These conditions correspond to the blue regions in Fig. \ref{fig:2}. In these regions, the information machine pumps the molecules from the low chemical potential reservoir to the high chemical potential reservoir, at the expense of increasing the entropy of the bit stream (writing information to the bit stream).

The information machine can also operate as an information eraser when the chemical potential difference is the predominant driving force, highlighted in Fig. \ref{fig:2} with red. When \(\delta>0\), namely there are more 0's than 1's in the incoming bit stream, the information will be (partially) erased when \(p'_0>p_0\) (more 0 is written) or \(p'_0<p_1\) (more 1 is written), which is equivalent to  \(\epsilon>\delta\) or \(\epsilon<-[3+K(\tau)]\delta/[3-K(\tau)]\). When \(\delta<0\), the information erasing conditions will be \(\epsilon<\delta\) or \(\epsilon>-[3+K(\tau)]\delta/[3-K(\tau)]\). Working in these regions of parameter space, the information machine consumes the Gibbs free energy to reduce the information entropy of the bit stream. Different interaction time periods \(\tau\) correspond to different periodic steady state solutions (\ref{eq:9}-\ref{eq:10}), and hence different boundaries in the phase diagram (see Fig. \ref{fig:2}). When \(\tau\rightarrow\infty\), the boundary of the ``eraser phase" will eventually approach the line \(\epsilon=-\delta\).

Our model thus reflects the competition between two effective driving forces. The operation regime is determined by which driving force is predominant. In Fig. \ref{fig:2}, one axis of the phase diagram is associated with the randomness of the incoming bits ($\delta$) and the other with the chemical potential difference ($\epsilon$). When the information machine is operated in the shaded area, either molecules are pumped against the chemical potential gradient, at the cost of simultaneously writing information to the bit stream; or the information encoded in the incoming bits is (partially) erased, at the expense of Gibbs free energy. The separating line between these two regimes is given by \(\delta=\epsilon\), where the two driving forces are comparable. In the unshaded regions, the information machine consumes both resources and is a dud.
\subsection{efficiency and power as functions of the parameters}
\begin{figure}
\centering
\includegraphics[width=.40\textwidth]{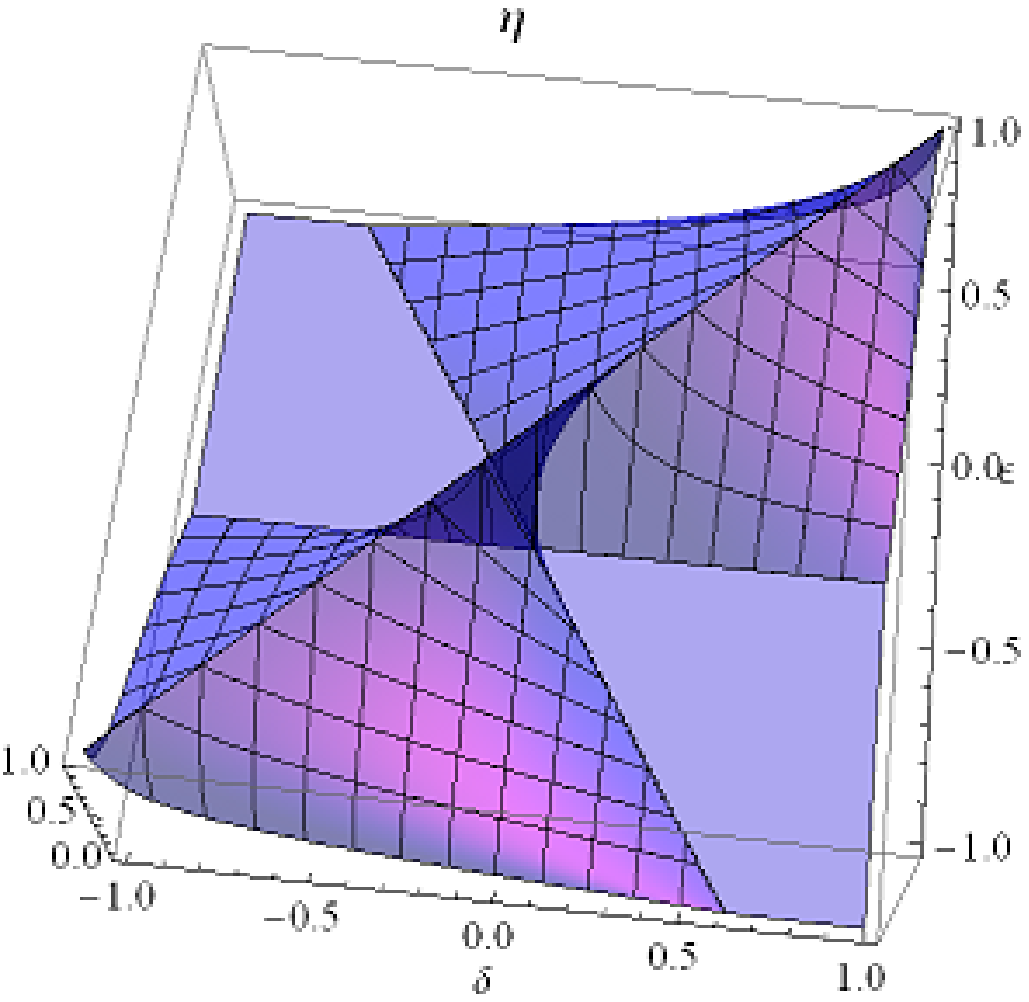}
\includegraphics[width=.45\textwidth]{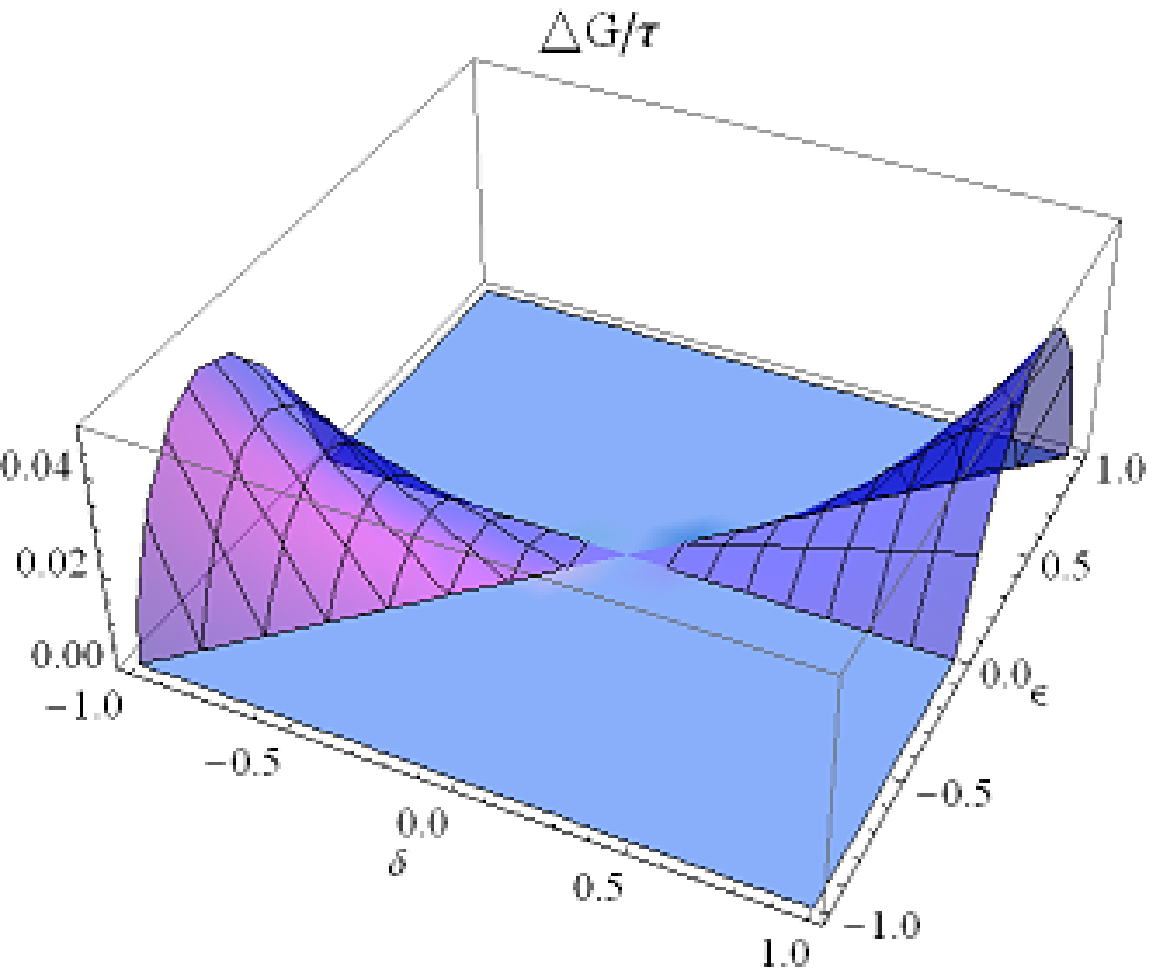}
\caption{Efficiency $\eta$ and power $\Delta G/\tau$ of the information pump and information eraser for $\tau=5$ and $\beta=1$.}
\label{fig:3}
\end{figure}

Besides determining the operation regimes of the information machine, we can further study the efficiency and the power of the information machine for an arbitrary given set of parameters $(\delta,\epsilon,\tau)$. \(\Delta G\) (\ref{eq:6}) is the free energy gained by the information machine during one period. The efficiency of the information pump can be expressed as
\begin{equation}\label{eq:11}
\eta(\delta,\epsilon;\tau)=\frac{\Delta G}{\ln{2} k_B T \Delta I},
\end{equation}
which approaches unity when \(\delta \to \epsilon\), but in this case the power is vanishingly small (see Fig. \ref{fig:3}). This result can be understood intuitively. When \(\delta \to \epsilon\), the two driving forces are almost equal to each other, so the process is very close to equilibrium. In another word, the irreversible entropy production is vanishingly small. Hence, the efficiency approaches its maximum value -- unity. In the information eraser regime, one can define the efficiency as
\begin{equation}\label{eq:11}
\eta'(\delta,\epsilon;\tau)=\frac{\ln{2} k_B T \Delta I}{\Delta G},
\end{equation}
and one can easily find that $\eta'\to 1$ when \(\delta \to \epsilon\) for the same reason (see Fig. \ref{fig:3}).

The power of the information pump $\Delta G/\tau$ is also illustrated in Fig. \ref{fig:3} (we do not study the power of the information eraser). It can be seen that the maximum power is reached when the driving force of the information reaches its maximum value $\delta=\pm 1$ and when the chemical potential difference is in the intermediate regime. This result can be understood intuitively: When the chemical potential difference $\Delta \mu$ is very large, the average number of molecules pumped in every period $\Phi$ is small. Hence, the power is low. When $\Delta \mu$ is very small, the average number of molecules pumped in every period is large, but the free energy gaining $\Delta \mu$ associated with one particle is very small. Hence, in both cases the power does not reach the maximum value.
\subsection{the second law}
Similar to the information engine \cite{Mandal2012} and the information refrigerator \cite{Mandal2013}, one can prove that the following inequality is satisfied in our model
\begin{equation}\label{eq:12}
\Delta G\leq \ln{2} k_B T \Delta I
\end{equation}
for any given set of (\(\epsilon,\delta,\tau\)). The equality holds only when \(\epsilon=\delta\) (see the SI in Ref.~\cite{Mandal2012}). This is also the upper limit of the efficiency for the information machine to pump molecules against the chemical potential gradient. When the information machine operates as an eraser, \(I(\delta')<I(\delta)\), the inequality is \(|\Delta G|\geq \ln {2}k_B T |\Delta I|\). This is also the upper limit of the efficiency to erase information. Both inequalities imply the Landauer's principle \cite{Landauer1961}. The entropy decrease of the reservoir, including {\it S} and {\it P} constituents as well as the heat bath, due to molecules transformation can be defined as \(\Delta I_r \equiv -\Delta G/\ln{2}k_B T\), and it must not exceed the entropy increase in the bit stream:
\begin{equation}\label{eq:13}
\Delta I_r+\Delta I\geq 0.
\end{equation}
If we identify the information entropy of the bit steam with the thermodynamic entropy, this inequality (\ref{eq:13}) indicates that the second law of thermodynamics is conserved in our autonomous information machine \cite{Mandal2012,Mandal2013}, and the Landauer's principle is equivalent to other formulations of the second law.
\section{Discussions and Summary}
In this article we propose a model of information pump and information eraser. This model provides a simple paradigm for the investigation of the thermodynamics of information processing based on enzyme kinetics.

\subsection{difference between the usual thermal engines and stochastic thermal machines}
The original development of thermodynamics found its inspiration in the steam engine. Nowadays the field is driven by the tiny molecular engines within living cells. Though of vastly differing scales, these engines share a common function: they transform energy into motion. For instance, ATP molecules provide the fuel for myosin molecules in muscle tissue to move along actin filaments, pulling the muscle fibers to which they are attached. Other motors are powered by light, by differences in proton concentrations or by differences in temperature \cite{Astumian2001a,Rubi2008}. Obviously the two kinds of engines also have different features (see e.g., section 9.6 of \cite{Kurzynski2006}). In the usual thermodynamic machines, such as a Carnot engine based on the classical ideal gas, the working substance for the thermodynamic machine is macroscopic in size, and usually the ensemble average of the thermodynamic variables is adequate to describe the system and the thermal fluctuations are usually ignored. In order to convert one form of energy into another form, the system must undergo certain thermodynamic cycles. However, in a stochastic thermal machine, for example in the Feynman's ratchet and Pawl \cite{Jarzynski1999}, Butticker-Landauer's ratchet \cite{Butticker1987,Landauer1988}, or in the information engine \cite{Mandal2012}, there are no thermodynamic cycles. They usually operate in (periodic) steady states. In addition, the thermal fluctuations are essential for the operation of the stochastic thermal machines, which puts a constraint on the upper limit of the size of the stochastic machines -- it must be microscopic in size so that it can be driven by the thermal fluctuations. 

\subsection{competing driving forces and stochastic thermal machines}
\begin{figure}
\centering
\includegraphics[width=.48\textwidth]{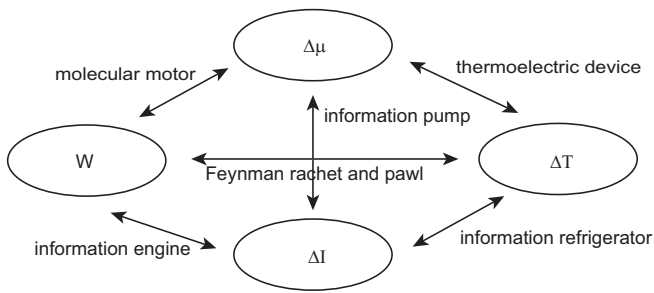}
\caption{Stochastic thermal machines and four kinds of thermodynamic driving forces: mechanical work $W$, chemical potential difference $\Delta\mu$, temperature difference $\Delta T$, and information $\Delta I$. Different stochastic thermal machines can be constructed between any pair of these thermodynamic driving forces. }
\label{fig_intro}
\end{figure}

Generally speaking, a stochastic thermal machine is a special kind of system that rectifies thermal fluctuations and converts one form of disordered energy into another form of ordered energy, by consuming certain kind of ordered resource, such as the information or the free energy. Two resources, or driving forces, compete with each other, generating one continuously in the cost of consuming another one. The driving forces can be the information, the temperature difference, the mechanical work, and the chemical potential difference. Fig. \ref{fig_intro} summarizes several known stochastic thermal machines. Besides the three information machines we mentioned in table \ref{table1}, there are three kinds of other stochastic thermal machines:  (1) the molecular motor which converts chemical potential difference, such as Gibbs free energy from ATP hydrolysis \cite{Austumian1996,LauLacoste-3,qian1997,qian2000}, into mechanical work, in the form of pulling loads along microtubes; (2) Feynman's rachet and pawl~\cite{Jarzynski1999} in which temperature gradient competes with the mechanical work; (3) thermoelectric devices \cite{Oisalvo1999} in which the chemical potential difference (voltage) and the temperature difference compete with each other ( Seeback effect and Peltier effect).

\subsection{information machines, feedback control and Maxwell's demon }
Among the stochastic thermal machines (Fig. \ref{fig_intro}), the information pump (current study), the information engine \cite{Mandal2012} and the information refrigerator \cite{Mandal2013} form a category of information machines (see Table \ref{table1}). They rectify thermal fluctuations and convert the disordered energy into other forms of ordered energy, such as the mechanical work by simultaneously consuming the information. It is worth mentioning that, they are different from a class of models with feedback control \cite{Sagawa2010} where the mutual information (or correlations) between the system and the measurement apparatus is identified as the information content. The mutual information can be harnessed to extract the ordered energy by rectifying the thermal fluctuations. In the information machine models, however, the correlations between the three-state system and the bits are eliminated artificially at the end of every period. Also, the information machines differ from the usual MD-assisted thermodynamic machines in the following aspects \cite{quan2006}: In the former cases it is impossible to identify the measurement process and the feedback control process. All processes are autonomous and simultaneous. Thus, the intelligent being is unnecessary. In the latter cases, however, there must be an intelligent being (or an external controller), so that the actions of the MD, such as the measurement and the feedback control, can be carried out.
\begin{table*}[tbp]
\caption{Three types of information machines: information engine \cite{Mandal2012}, information refrigerator \cite{Mandal2013} and information pump.}
\begin{center}
\begin{tabular}{c|c|c}
\hline
Information Machines &Input& Output\\ \hline\hline
Information Engine \cite{Mandal2012} &Information & Mechanical work\\
Information Refrigerator \cite{Mandal2013} &Information& Heat flow against temperature gradient\\
Information Pump &Information& Molecule flux against chemical potential gradient\\ \hline\hline
\end{tabular}
\end{center}
\label{table1}
\end{table*}

\subsection{molecular pumps in biological systems}

In the past two decades, the thermodynamic machinery of life \cite{Kurzynski2006} has attracted a lot of attention, and studies in this field bloomed in recent years \cite{LanSartori-5,Qian-69,MehtaSchwab-70,SkogeNaqvi-71,QianBeard-22,Seifert2011,Andrieux2008,ItoSagawa-5,Andrieux2013,Seifert2012,Jordan2014}. Actually, in biological systems, instead of the temperature gradient, the chemical potential is a prevailing driving force \cite{Jarzynski2015}. Living systems consume the Gibbs free energy to perform physical and chemical functions \cite{Kurzynski2006,QianBeard-22,EisenbergHill-4}. The chemical potential difference comes mainly from ATP hydrolysis, or ion transport. 

In our model, the enzyme is a stochastic machine and the enzyme reactions are naturally built on the molecular level. The enzyme system \cite{Seifert2011} is less artificial than the mechanical models \cite{Mandal2012,Mandal2013,lu2014engineering} and has been studied extensively. The experimental techniques accumulated in the field of enzyme kinetics will facilitate the realization of the Maxwell's thought experiment. In fact, there are many cycle reactions that have net fluxes in cells, such as the phosphorylation-dephosphorylation cycle \cite{Qian-1} driven by ATP hydrolysis. As long as we can design reactions with asymmetric kinetics like the one designed in our model, it is possible to build an autonomous information pump that produces molecules with higher chemical potential, without consuming other free energies.


\subsection{concluding remarks}
In summary, in this article, we propose an autonomous information machine that can pump molecules against the chemical potential gradient without the expenditure of work or free energy, or (partially) erase information encoded in the bit stream. 
Information is a new driving force that can be used to pump molecules against the chemical potential gradient. We solve the dynamics exactly and obtain the ``phase diagram" in the parameter space. In certain regions of the parameter space the information machine operates as an information pump, while in some other regions of the parameter space it operates as an information eraser. In both regions, the Landauer's principle (one formulation of the second law of thermodynamics) is conserved. Our model offers a simple paradigm for the studies of thermodynamics of information processing by small systems. Our investigation not only brings important insights to the understanding about the intrinsic relations between information, free energy, entropy, and work, but also sheds new light on the ultimate realization of the Maxwell's thought experiment in the laboratory.


\acknowledgments
We thank Hao Ge, Hong Qian, C. Jarzynski, D. Mandal and Yuhai Tu for helpful discussions. H.T.Q. gratefully acknowledges support from the National Science Foundation of China under grant 11375012, and The Recruitment Program of Global Youth Experts of China.

\appendix

\section{solution to the information pump}
From Eq.~(\ref{sixstate}) and the assumptions in section III. A, the transition matrix can be written as:
\begin{equation}\mathcal{R}=\left(\begin{array}{cccccc}-1&1+\epsilon&0&0&0&0\\1&-2&1&0&0&0\\
0&1-\epsilon&-2&1&0&0\\0&0&1&-2&1+\epsilon&0\\0&0&0&1&-2&1\\0&0&0&0&1-\epsilon&-1\\
\end{array}\right).\end{equation}
We find that the matrix $\mathcal{R}$ can be decomposed into $\mathcal{R}=UN^{-1}\Lambda V$. Here $\Lambda=\mathrm{diag}(0,-c,-1,-2,-3,-d)$ and $N=\mathrm{diag}(6,12c,4,6,12,12d)$, which are exactly the same as those in Ref.~\cite{Mandal2012}
while $U$ and $V$ are different:
\begin{subequations}
\begin{equation}U=\left(\begin{array}{cccccc} xy^{-1}&x&1&x&xy^{-1}&x\\y^{-1}&-a&0&-1&-2y^{-1}&-b\\1&a+y&-1&-x&1&b+y\\1&-a-x&-1&y&1&-b-x\\
x^{-1}&a&0&1&-2x^{-1}&b\\x^{-1}y&-y&1&-y&x^{-1}y&-y
\end{array}\right),\end{equation}
\begin{equation}V=\left(\begin{array}{cccccc}xy&xy&xy&xy&xy&xy\\y&-ay&a+y&-(a+x)&ax&-x\\y^2&0&-xy&-xy&0&x^2\\
y&-y&-x&y&x&-x\\xy&-2xy&xy&xy&-2xy&xy\\y&-by&b+y&-(b+x)&bx&-x\\
\end{array}\right).\end{equation}
\end{subequations}
Here $a=1-\sqrt{3},b=1+\sqrt{3},x=1+\epsilon,y=1-\epsilon,c=2-\sqrt{3},d=2+\sqrt{3}$.

Following the same procedure as that in Ref.~\cite{Mandal2012} we find the periodic steady state $\mathbf{q}^{pss}$ by solving
\begin{equation}\mathcal{T}\mathbf{q}^{pss}=\mathbf{q}^{pss},\end{equation}
where $\mathcal{T}=\mathcal{P}_De^{\mathcal{R}\tau}\mathcal{M}=\mathcal{P}_DUN^{-1}e^{\Lambda\tau}V\mathcal{M}$, and $\mathcal{P}_D=(\mathcal{I},\mathcal{I}), \mathcal{M}=(p_0\mathcal{I},p_1\mathcal{I})^T$, with $\mathcal{I}$ to be $3\times3$ identical matrix. $p_0$ and $p_1$ are determined by $\delta=p_0-p_1$ and $1=p_0+p_1$. After a straightforward calculation we obtain:
\begin{equation}
\begin{split}
\mathcal{T}=&\frac{1}{12}\left(\begin{array}{ccc}F+G+\delta H&M-2\delta L&F-G+\delta H\\M&M+12\sigma^3&M\\F-G-\delta H&M+2\delta L&F+G-\delta H\end{array}\right)\\
&+\frac{\epsilon}{12}\left(\begin{array}{ccc}F-G-H&M+2L&F-G-H\\0&0&0\\G+H-F&-M-2L&G+H-F\\ \end{array}\right),
\end{split}
\label{Gamma}
\end{equation}
with $\sigma=e^{-\tau}$, $F=4+2\sigma^3,G=4\sigma^2+\sigma^c+\sigma^d,H=\sqrt{3}(\sigma^c-\sigma^d),L=2\sigma^2-\sigma^c-\sigma^d$, and $M=4-4\sigma^3$. Based on Eq.~(\ref{Gamma}) we obtain:
\begin{equation}\mathbf{q}^{pss}=\frac{1}{3}\left(\begin{array}{c}1+E\\1\\1-E\end{array}\right), E=\frac{(H-L)(\delta-\epsilon)}{6-G}+\epsilon.\end{equation}

Now we can calculate $\Phi$ defined by Eq.~(\ref{eq:5}). Since $\Phi=p_1'-p_1$, the only remaining problem is to solve $p_1'$. Actually we have \cite{Mandal2012}
\begin{equation}\left(\begin{array}{c}p_0'\\p_1'\end{array}\right)=\mathcal{P}_Be^{\mathcal{R}\tau}\mathcal{M}\mathbf{q}^{pss}
,\mathcal{P}_B=\left(\begin{array}{cccccc}1&1&1&0&0&0\\0&0&0&1&1&1\\\end{array}\right).\end{equation}
A direct calculation gives:
\begin{subequations}
\begin{equation}
\Phi (\delta,\epsilon;\tau)=\frac{\delta-\epsilon}{2}[1-\frac{1}{3}K(\tau)],
\end{equation}
where
\begin{equation}
K(\tau)=e^{-2\tau}\frac{(1+8\alpha+4\sqrt{3}\beta)-(2+7\alpha+4\sqrt{3}\beta)e^{-2\tau}}{3-(2+\alpha)e^{-2\tau}}.
\end{equation}
\end{subequations}

\section{The solution to a general transition matrix}
Consider a general class of models with the transition matrix (\ref{sixstate}):
\begin{equation}
\mathcal{R}=\left(\begin{array}{cccccc}-b_1&a_1&0&0&0&0\\b_1&-a_1-b_3&a_3&0&0&0\\0&b_3&-a_3-b_2&a_2&0&0\\0&0&b_2&-a_2-b_1&a_1&0\\0&0&0&b_1&-a_1-b_3&a_3\\0&0&0&0&b_3&-a_3\end{array}\right).
\label{B1}
\end{equation}
We will give a general spectral decomposition of the matrix in form of
\begin{equation*}
\begin{split}
\mathcal{R}=UN^{-1}\Lambda V=&\left(\begin{array}{ccc}\uparrow&&\uparrow\\ \mathbf{u_1}&\cdots&\mathbf{u_6}\\ \downarrow &&\downarrow\end{array}\right)\left(\begin{array}{ccc}n_1^{-1}&&\\&\ddots&\\&&n_6^{-1}\end{array}\right)\\
&\times \left(\begin{array}{ccc}\lambda_1&&\\&\ddots&\\&&\lambda_6\end{array}\right)\left(\begin{array}
{ccc}\leftarrow&\mathbf{v}_1&\rightarrow\\ &\vdots& \\ \leftarrow&\mathbf{v}_6&\rightarrow\end{array}\right),
\end{split}
\end{equation*}
where $a_i$ and $b_i$ ($i=1,2,3$) are all real positive numbers. The rows of $V$ are the left eigenvectors, and the columns of $U$ are the right eigenvectors, and $VU=N$.  The eigenvalues are
\begin{equation}\{0,\omega_1,v_1,\omega_2,v_2,\omega_3\},\end{equation}
where $v_{1,2}$ are the roots of equation
\begin{equation}(\lambda+c_1)(\lambda+c_3)-a_1b_3=0,\label{B4}\end{equation}
with $c_i=a_i+b_i (i=1,2,3)$ ; $\omega_{1,2,3}$ are the roots of equation
\begin{equation}
\begin{split}
(\lambda+c_1)(\lambda+c_2)(\lambda+c_3)&-a_1b_3(\lambda+c_2)-a_2b_1(\lambda+c_3)\\
&-a_3b_2(\lambda+c_1)=0.\label{B5}
\end{split}
\end{equation}
It can be proved that $v_1,v_2,\omega_1,\omega_2,\omega_3 $ are all nonzero (0 is obviously not a root of Eq.~(\ref{B4}) or Eq.~(\ref{B5})) real numbers. They differ from each other, so we don't have to worry about the cases with degenerate or complex roots \footnote{From Eq.~(\ref{B4}) we know that $\Delta = (c_1-c_3 )^2 + 4a_1 b_3>0$, so that $v_1 \neq v_2$ and are both real. If we define $f(\lambda) \equiv $l.h.s of Eq.~(\ref{B5}) and $c_m \equiv \min \{ c_1,c_2,c_3 \}$, $ c_M \equiv \max \{ c_1,c_2,c_3 \}$, if $c_m < c_M$, we will have $f(-c_M) > 0 > f(-c_m)$, thus $f(\lambda) = 0$ have three different real roots respectively among the intervals $(-\infty,-c_M)$, $(-c_M,-c_m)$ and $(-c_m,+\infty)$. Otherwise $c_m = c_M = c$, the three roots can be straightforwardly evaluated as $-c$, $-c \pm \sqrt{d}$, where $d \equiv a_1 b_3 + a_2 b_1 + a_3 b_2$. Finally, we reversely assume that $\lambda_0 \in \mathbb{R}$ is a mutual root of Eq.~(\ref{B4}) and Eq.~(\ref{B5}), after some algebraic transformations we obtain an equality $a_2 b_1 (\lambda_0 + c_3 )^2 + a_1 b_2 a_3 b_3 = 0$, which can never hold, hence the contradiction implies $\{ \omega_1,\omega_2,\omega_3 \} \cap \{ v_1,v_2 \} = \emptyset$.}. The corresponding right eigenvectors are
\begin{widetext}
\begin{equation*}
\left(\begin{array}{ccc}\uparrow&&\uparrow\\ \mathbf{u_1}&\cdots&\mathbf{u_6}\\ \downarrow &&\downarrow\end{array}\right)=U
=\left(\begin{array}{cccccc}\frac{a_1a_3}{b_1b_3}a_2&a_1a_3&\frac{v_1+c_3}{b_3}a_2&a_1a_3&-\frac{v_2+c_3}{b_3}a_2&a_1a_3\\
\frac{a_3}{b_3}a_2&a_3(b_1+\omega_1)&-\frac{v_1+a_3}{b_3}a_2&a_3(b_1+\omega_2)&\frac{v_2+a_3}{b_3}a_2&a_3(b_1+\omega_3)\\
a_2&F(\omega_1)&-a_2&F(\omega_2)&a_2&F(\omega_3)\\b_2&-G(\omega_1)&-b_2&-G(\omega_2)&b_2&-G(\omega_3)\\
\frac{b_1}{a_1}b_2&-b_1(a_3+\omega_1)&-\frac{v_1+b_1}{a_1}b_2&-b_1(a_3+\omega_2)&\frac{v_2+b_1}{a_1}b_2&-b_1(a_3+\omega_3)\\
\frac{b_1b_3}{a_1a_3}b_2&-b_1b_3&\frac{v_1+c_1}{a_1}b_2&-b_1b_3&-\frac{v_2+c_1}{a_1}b_2&-b_1b_3\\
\end{array}\right),
\end{equation*}
\end{widetext}
where $F(\lambda)=(\lambda+c_1)(\lambda+b_3)-a_1b_3, G(\lambda)=(\lambda+a_1)(\lambda+c_3)-a_1b_3$.
The left eigenvectors of $\mathcal{R}$ can be obtained by solving $\mathcal{R}^T\mathbf{v}_k^T=\lambda_k\mathbf{v}_k^T, (k=1,2,\cdots,6)$ in the manner of the conventional right eigenvector problem. The results are as follows:
\begin{widetext}
\begin{equation}
\left(\begin{array}{ccc}\leftarrow&\mathbf{v}_1&\rightarrow\\ &\vdots& \\ \leftarrow&\mathbf{v}_6&\rightarrow\end{array}\right)=V=
\left(\begin{array}{cccccc}1&1&1&1&1&1\\ \frac{b_2}{a_1a_3}&\frac{b_1+\omega_1}{a_1a_3b_1}b_2&\frac{F(\omega_1)}{a_1a_3b_1b_3}b_2&-\frac{G(\omega_1)}{a_1a_3b_1b_3}a_2&-\frac{a_3+\omega_1}{b_1b_3a_3}a_2&-\frac{a_2}{b_1b_3}\\
\frac{b_1}{a_1a_3}(v_1+c_3)&-\frac{v_1+a_3}{a_3}&-1&-1&-\frac{v_1+b_1}{b_1}&\frac{a_3}{b_1b_3}(v_1+c_1)\\
\frac{b_2}{a_1a_3}&\frac{b_1+\omega_2}{a_1a_3b_1}b_2&\frac{F(\omega_2)}{a_1a_3b_1b_3}b_2&-\frac{G(\omega_2)}{a_1a_3b_1b_3}a_2&-\frac{a_3+\omega_2}{b_1b_3a_3}a_2&-\frac{a_2}{b_1b_3}\\
-\frac{b_1}{a_1a_3}(v_2+c_3)&\frac{v_2+a_3}{a_3}&1&1&\frac{v_2+b_1}{b_1}&-\frac{a_3}{b_1b_3}(v_2+c_1)\\
\frac{b_2}{a_1a_3}&\frac{b_1+\omega_3}{a_1a_3b_1}b_2&\frac{F(\omega_3)}{a_1a_3b_1b_3}b_2&-\frac{G(\omega_3)}{a_1a_3b_1b_3}a_2&-\frac{a_3+\omega_3}{b_1b_3a_3}a_2&-\frac{a_2}{b_1b_3}\\
\end{array}\right).
\end{equation}
\end{widetext}
Then we calculate $n_k=\mathbf{v}_k^T\mathbf{u}_k$:
\begin{equation}
\begin{split}
n_1=&(a_1a_3+a_3b_1+b_1b_3)(\frac{a_2}{b_1b_3}+\frac{b_2}{a_1a_3}),\\
\end{split}
\end{equation}
\begin{equation}
\begin{split}
n_{2i}=&b_2[1+\frac{(b_1+\omega_1)^2}{a_1b_1}+\frac{F^2(\omega_i)}{a_1a_3b_1b_3}]\\
&+a_2[1+\frac{(a_3+\omega_i)^2}{a_3b_3}+\frac{G^2(\omega_i)}{a_1a_3b_1b_3}],\\
\end{split}
\end{equation}
\begin{equation}
\begin{split}
n_{2j+1}=&-\frac{a_2v_j}{a_1a_3b_3}[(c_3+v_j)^2+a_1b_3]\\
&-\frac{b_2v_j}{a_1b_1b_3}[(c_1+v_j)^2+a_1b_3],
\end{split}
\end{equation}
where $i=1,2,3$ and $j=1,2$. The transition matrix in Ref. \cite{Mandal2012} is a special case of Eq.~(\ref{B1}) when $a_1=a_3=b_1=b_3=1, a_2=1+\epsilon, b_2=1-\epsilon, c_i=2$ $(i=1,2,3)$. While the transition matrix (\ref{sixstate}) in our current investigation corresponds to another special case with with $a_1=1+\epsilon,b_3=1-\epsilon,b_1=a_2=b_2=a_3$, i.e., $c_1=2+\epsilon,c_1=2,c_1=2-\epsilon$.

\bibliography{CJ_references}

\end{document}